\documentclass[3p,times]{elsarticle}
\usepackage{ecrc}
\usepackage{graphicx}
\usepackage{color}
\usepackage{dcolumn}
\usepackage{bm}
\usepackage{relsize}
\newcommand{\be}{\begin{equation}}
\newcommand{\ee}{\end{equation}}
\newcommand{\ba}{\begin{eqnarray}}
\newcommand{\ea}{\end{eqnarray}}
\newcommand{\bd}{\begin{displaymath}}
\newcommand{\ed}{\end{displaymath}}

\volume{00}
\firstpage{1}
\journalname{Physica A}
\runauth{Csernai et al.}
\jnltitlelogo{Physica A}

\begin{document}
\begin{frontmatter}

\title{Quantitative assessment of increasing complexity}
\author{L.P. Csernai, S.F. Spinnangr, S. Velle}
\address{ Institute of Physics and Technology, 
University of Bergen, Allegaten 55, 5007 Bergen, Norway }

\date{\today}
\begin{abstract}
We study the build up of complexity on the example of 1 kg matter in different
forms. We start with the simplest example of ideal gases, and then continue
with more complex chemical, biological, life, social and technical 
structures. We assess the complexity of these systems quantitatively, 
based on their entropy. We present a method to attribute the same 
entropy to known physical systems and to complex organic molecules,
up to a DNA molecule. The important steps in this program and the 
basic obstacles are discussed.
\end{abstract}
\end{frontmatter}


\section{Introduction}

The problem of development is longstanding in humanity. It became 
quantitative with the development of statistical physics and
quantum physics. The first theoretical step was done by Boltzmann,
who laid down the basis of microscopic quantitative treatment of
entropy both for equilibrated and out of equilibrium systems. In the
H-theorem he showed that closed systems in equilibrium maximize 
their entropy, and all microscopic interactions drive the system
towards this equilibrium. This macro state is the most probable one
the system can reach.

Consequently, a less probable, non-equilibrium state can be formed only
if the system is not closed and can exchange entropy with the 
surrounding, so that its own entropy decreases. Under stationary
conditions, constant pressure and temperature, spontaneous chemical
reactions do not lead to more complex systems.

These ideas were discussed already by E. Schr\"odinger in 1943,
in his book: "What is life?"
\cite{Schroedinger}.
He described that life forms are 
highly complex systems of a high level of "order", that can develop from
disordered systems, or more probably from ordered systems of a
somewhat lower level of order. 
Our ultimate aim is to show at what cost sustainable development
is possible, what the direction of sustainable development is, and
which processes are working towards such development. We aim to quantify 
the LEVEL and the RATE of the development quantitatively.

Schr\"odinger described the problem and the concepts in a genius
way, but he could not give quantitative information at that time. We know 
much more about biological structures and life forms today, so
it is possible to discuss these problems quantitatively.

In this work we introduce the method to calculate the entropy of a well 
defined amount of matter, irrespective of what form of matter we are discussing.
This is because the development on the Earth happens with a constant amount of 
matter, but the forms of matter may change due to nuclear, chemical, biological,
technological, intellectual and societal reactions and changes.  Thus, 
we introduce a unit of {\bf 1 kg} for our discussion.  (Although, for 
intellectual and societal changes and structures this choice of unit is 
too extreme in this moment.)  The human intellect is 
the limit where the material form, the nervous system,
and the intellectual information content are at the boundary of our knowledge.
At this time this program can be performed up to simple structures of actions 
and simple vegetative nervous systems. For transparency and the illustration 
of the program we start from  the simplest (and highest entropy) systems.  
We intend to provide a simple and sometimes even simplified presentation of 
the entropy calculations and the corresponding degrees of freedom of materials.

The second important new aspect we introduce, is to achieve the possibility of
comparing different forms of matter with a uniform definition of entropy. In 
the present literature, in mathematics, information theory, linguistics, etc. 
many definitions of entropy exist. The most widely known, after the material 
entropy in statistical physics, is the "Shannon entropy'' \cite{Shannon}. 
We remind the 
reader that this definition was already used by Boltzmann in the H-theorem 
for non-equilibrium systems, and this makes it possible to attribute the same 
physical dimension to both entropies.

A third aspect we have to utilize is the quantization of the phase space entropy
by the volume of the phase-space cell. This is an important step in unifying 
the entropy evaluation and it is not obvious how to perform this for highly 
complex systems. 

The last step is to select the physically realized system configuration(s) 
from all possible ones with the same degree of complexity and degrees of 
freedom. In highly complex systems this step is also nontrivial, and for 
societal or intellectual systems it is even debatable which systems are 
realized or realizable.

We introduce this system in a series of examples, from the most simple 
elementary ones up to the most complex ones, which can still be discussed 
on a quantitative basis. These last examples are from biology, but our aim 
is not to contribute to quantitative biology, but rather to demonstrate our 
procedure, how to quantitatively assess the sustainable development 
following Schr\"odinger's  original ideas.

\begin{table}[h]   
\begin{center}
\begin{tabular}{crrrr} \hline\hline \phantom{\Large $^|_|$}
Material\ \  &  $A_P$ & $ N_P $ & $\sigma_P$ & $S_{1kg}$ \\
    &    & (mol/kg) &  &  (J/K)  \\
\hline
$H_2$ & 2   & 496.046  & 14.146 & 58344.0 \\
$He$  & 4   & 248.023  & 15.186 & 31316.1 \\
$H_2O$& 18  & 55.116   & 17.442 & 7993.0  \\
$H_2O^*$& 18  & 55.116   & 17.988 & 8243.0  \\
$Rn$  & 222 & 4.469    & 21.211 & 788.1   \\
$WF_6$& 298 & 3.329    & 21.652 & 599.3   \\
$UF_6$& 352 & 2.818    & 21.902 & 513.2   \\
$C60$ & 720 & 1.378    & 22.975 & 263.2   \\
\hline
\end{tabular}
\end{center}
\caption{
Thermodynamical parameters of 1kg material in
different forms approximated as ideal gases, depending on their mass 
numbers, $A_P$, different mol-numbers of particles, $N_P$, and this 
number is decreasing with increasing $A_P$. The specific entropy
of the composite particles is indicated by the dimensionless  $\sigma_P$
\ \ ($\hbar, c, k_B = 1$).  The total entropy of
the material at the used $T=300^o$K and $p=1$bar, 
is also decreasing with increasing
complexity, i.e. increasing $A_P$.
(For $H_2O^*$ the temperature is taken to be $T=100^o$C$=373.15^o$K.)
}
\label{t1}
\end{table}   

\section{Elementary Entropy Evaluations for Unit Amount of Matter}
\label{s0}

In this section we introduce the first steps of our program, with the basic 
units and definitions using the most simple, generally known systems.  
For those who are well familiar with statistical physics the sections
up to Section 4 (ref\{S3\}) or 5 (ref\{S4\}) can be skimmed through.
Here the importance of quantization and the unified treatment of Gibbs and
Shannon entropy are essential.

We can start with the example of more and more complex chemical
structures. For example with dilute gases, approximated first as 
ideal gases.

We take a series of gases, $H_2$, $He$, $H_2O$, $Rn$, $WF_6$, $UF_6$
and $C_{60}$, with increasing molecular weight or "Particle mass Number",
$A_P$.  We take $1$kg of material, so the number of particles, $N_P$, in
this amount of matter will decrease with increasing mass number, $A_P$.
We choose that the gas is at standard atmospheric pressure,  
i.e. at $P=1$atm, and $T=300^o$K temperature.

Using the ideal gas approximation
\be
S_{1{\rm kg}} = 
N_P \, k_B \left[ \frac{5}{2}
+ \ln \left(\frac{(2 \pi m_P c^2 k_B T)^{3/2}}{n_P(2\pi\hbar c)^3}\right)
	            \right]\ ,
\label{eq1}		
\ee
where $k_B$ is Boltzmann constant 
($k_B=1.38064852 \cdot 10^{-23}$J/K), $m_P$ is the particle mass, 
(we take it as $m_P=A_P m_{Nucl.}$
in terms of the nucleon mass, $m_{Nucl.}$) and $n_P$ is the particle
density $(n_P = Av / (V_{I.G.} )$ in terms of the Avogadro number,
$Av$. The molar volume of an ideal gas is, $V_{I.G.}$ at STP, standard pressure
and standard temperature of $T=273.15^o$K).
The moderately increasing dimensionless specific entropy per particle,
$\sigma_P=S/(k_B\, N_P)$, is also shown.
 See Table \ref{t1}. 
In the table all values were calculated in the ideal gas
 approximation and for $T=300^o$K, except the water, $H_2O$, 
 which is at boiling temperature, $T=373.15^o$K.

Noble gases, $He$, $Rn$, and also small molecules, can be well 
approximated as ideal gases, their vibrational degree of freedom
carries negligible energy.
Under standard conditions the heaviest couple of materials are
not gases and their interactions and internal degrees of freedom 
should also be taken into account. Thus, for water vapour and for the heavy
gases the ideal gas approximation underestimates the entropy of the
material.

\begin{figure}[ht]     
\begin{center}
\resizebox{0.6\columnwidth}{!}
{\includegraphics{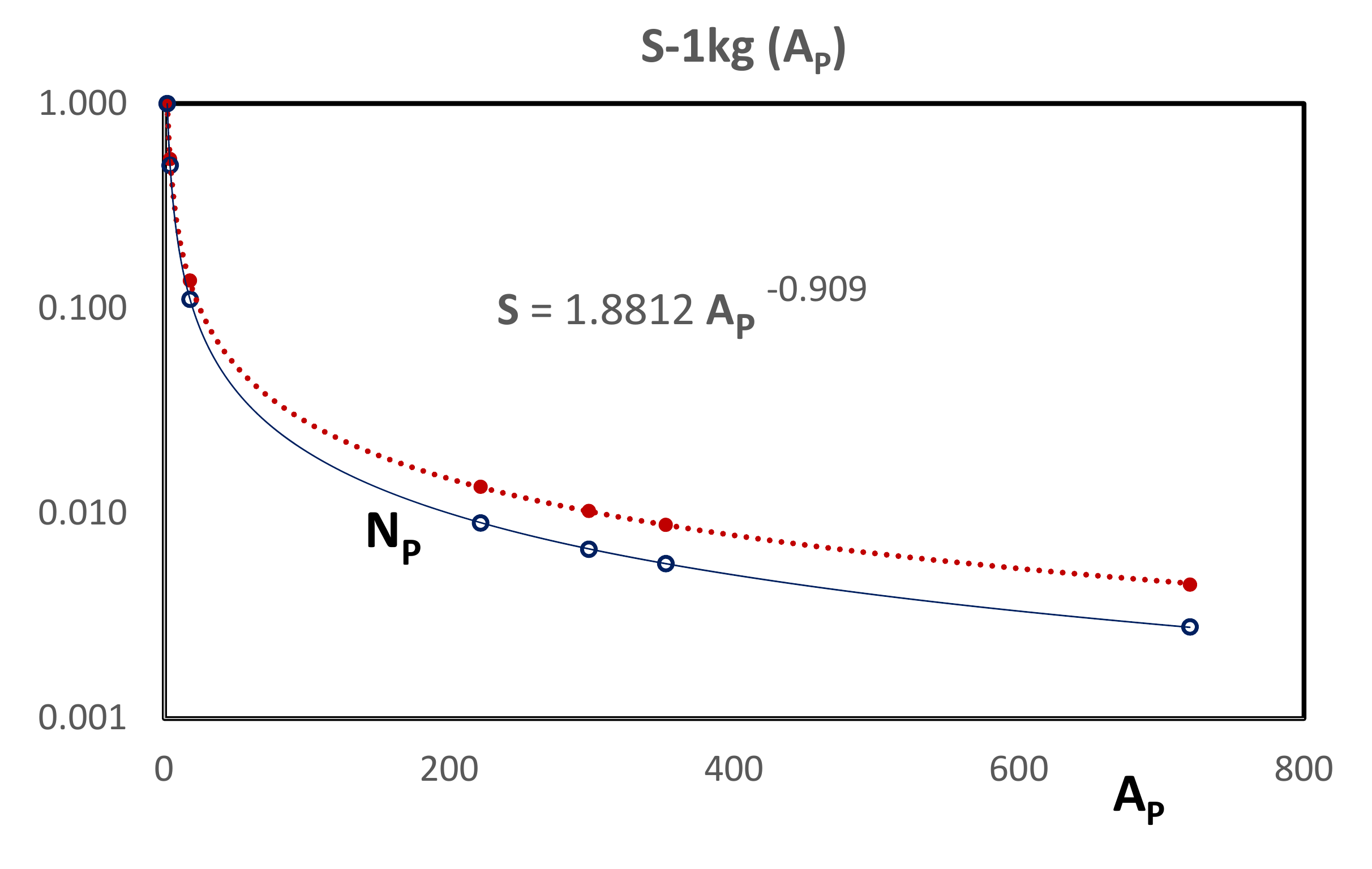}}
\caption{
(color online)
The mass number, $A_P$, dependence 
of the relative entropy of 1kg material, $S$, compared to
the number of these particles, $N_P$.
The relative entropy, $S$, as well as the particle number, $N_P$,
of 1 kg $H_2$ is taken to be unity for the comparison.
The entropy decreases slower than the decrease of the
particle number. This means that the entropy per 
particle is increasing.
}
\label{f1}
\end{center}
\end{figure}

Nevertheless, even for ideal gases the entropy is not behaving 
the same way as the particle number. This can be seen well in
Fig. \ref{f1}. Let us take the values of $N_P$ and $S_P(1kg)$
for $H_2$
as standard unity, and see how $N_P$ and $S_P(1kg)$ change with
increasing $A_P$. While $N_P$ decreases according to the 
$N_P A_P = $const. constraint (and therefore $N_P \propto A_P^{-1}$),
the entropy, $S_P(1kg)$, decreases 
less due to the additional contribution of the log term in the 
entropy expression $\propto \ln(m_P/n_P)$. Consequently the 
entropy decreases as $S \propto A_P^{-0.9}$.

%

\begin{figure}[h]     
\begin{center}
\resizebox{0.6\columnwidth}{!}
{\includegraphics{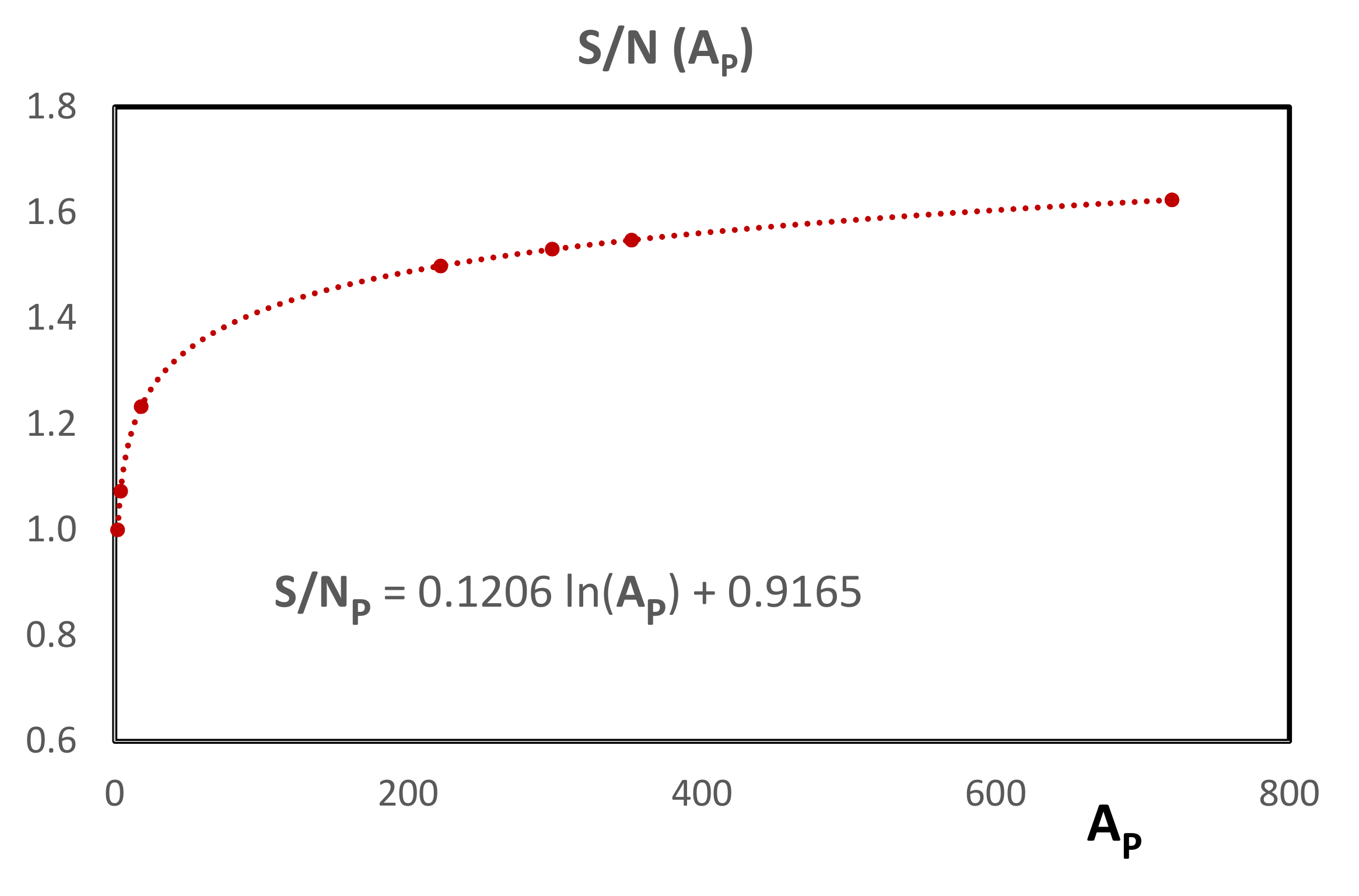}}
\caption{
(color online)
The change of the relative ratio of entropy versus particle number,
$S/N_P$ as a function of the mass number of a given particle 
species, $A_P$. For this comparison the $S/N_P$ for $H_2$
is taken to be unity. 
}
\label{f2}
\end{center}
\end{figure}

This change is shown if Fig. \ref{f2}, where again the ratio 
$\sigma_P=S/(k_B\, N_P)$ for $H_2$ is taken to be unity. The increasing 
relative entropy
exceeds $S/(k_B\, N_P) = 1.6$ for $C_{60}$ but it has a saturating
tendency. In this ideal gas approximation the ratio
is not expected to exceed two.

For larger, more complex molecules the number of degrees of freedom 
increases, and specific molecular configuration has less 
entropy than the completely random ideal gas.

Interacting materials can form other liquid or solid phases, which
have more constraints, compared to the increasing number of degrees
of freedom. This decreases the entropy of a given liquid or solid 
further. We will illustrate this on the example of water.

As in liquids there is still considerable room for random configurations
where entropy is decreasing less than that of the solids, where the
level of "order"
\cite{Schroedinger}
is higher.

Schr\"odinger's considerations were extended to lifeforms of
matter 
\cite{Schroedinger}, and similar comparative studies were used
for comparing the change of the entropy content of different
species during their lifespan
\cite{PCs1980},
as well as the rate of change of the entropy during the life of different
species. The initial development phase leads to a considerable 
decrease of the entropy of the matter incorporated into the 
living species (build up of neg-entropy), then the entropy is 
increasing again in the second stage of the life of the species.
The rate of these changes is connected to the energy exchange
with the environment (or metabolism) and this then determines 
the lifespan of a species.

Interestingly Schr\"odinger discusses two ways on how to achieve 
a high level of order, (i.e. smaller entropy).  High level of
order can be built up from dis-ordered materials as well as from
an already ordered material with achieving a higher level of order.
This second way is obviously easier and preferable.

Interestingly the same fundamental ideas can be extended to the
sustainable development of the Earth, and this has also been done 
for a while, using the principles of statistical physics
\cite{CsPSX2016}.
The consequences of these more fundamental and more quantitative
considerations are interesting. The popular folklore considers
some selected energetic processes as sustainable, while others
are not. Recently some energy resources are declared "renewable"
others are not. In this respect the use of "bio-fuels" is 
quite problematic, because if we burn or destroy
highly complex, biological 
materials, this may lead to decreasing "order" in 
Schr\"odinger's sense, which is not contributing to sustainable
development. The same issue may arise if photovoltaic production
sites occupy large agricultural territories, which is not 
taken into consideration.

Therefore our aim is to demonstrate the connection between
sustainable development, energy and entropy exchange in 
a quantitative way, as far as possible.

Furthermore, we want to extend these considerations beyond
different life-forms of matter, to technical and social
development. In this last point, quantitative energetic and
entropy aspects are probably beyond our present knowledge,
but one can still estimate which direction of changes
certain social actions can or will cause.

\section{Network or topological entropy}
\label{S2}

We can continue the previous simplified study by considering 
different hypothetical ideal gases constructed from nucleons
like $H_1$,  $H_2$,  $H_3$, $H_4$. If we consider these as different 
ideal gases without taking into account the type of binding, then
we end up with the result in the previous section. Let us neglect the
physical features of a binding, such as its energy or extra degrees of 
motion, and only consider the possible topologies of the binding.
If we take the molecule $H_2$, then we have a link between two nucleons.
For this molecule we can only insert a link one way to make a cluster,
so the existence of the link does not contribute to extra energy or
entropy (because we neglected the small rotational or vibrational energies).

In both cases ( $H_1$ and  $H_2$) there is only one configuration for the
molecule, i.e. $i=1$ and $p_1 = 1$, 
and the sum of all allowed configuration states is, $N$, is $N=1$.
Thus, the topological Shannon entropy of these molecules is
\be
H(X) = - p_1 \ln p_1 = 0 .
\label{e2}
\ee
If $N$ would be $N > 1$, then the most random configuration would
be $p_i = 1/N$ for each $i$ and so for this configuration the Shannon
entropy is
\be
H(X) = - \sum_{i=1}^N  p_i \ln p_i = 
- N  \frac{1}{N}  \ln \frac{1}{N} = \ln N = H_{max} .
\label{HX}
\ee

Let us now consider a (hypothetical)  $H_3$-molecule. This can be formed
by inserting (2) links or (3) links! Thus, we can have two characteristic 
structures for 3 nucleons. Two links, (2), can be inserted 3 ways: 12 \& 23 or
13 \& 23 or 13 \& 12 for identical but distinguishable nucleons. 
Three links, (3), can be inserted only one way, 123, for identical particles. 
The two link configurations can be obtained from the three link
ones by cutting one link and this can be done in three ways.
This is altogether 4 configurations, with (3) being $i=1$ and
(2) being $i=2$, then $p_1=1/4$ and $p_2=3/4$.
The Shannon entropy of a system $X$ with all possible 
configurations of the $H_3$ molecule 
 is then
\be
H(X) = 
- \left[  \frac{1}{4}  \ln \frac{1}{4} + 
\frac{3}{4}  \ln \frac{3}{4} \right]
= 0.5623 .
\ee

\begin{figure}[h]     
\begin{center}
\resizebox{0.3\columnwidth}{!}
{\includegraphics{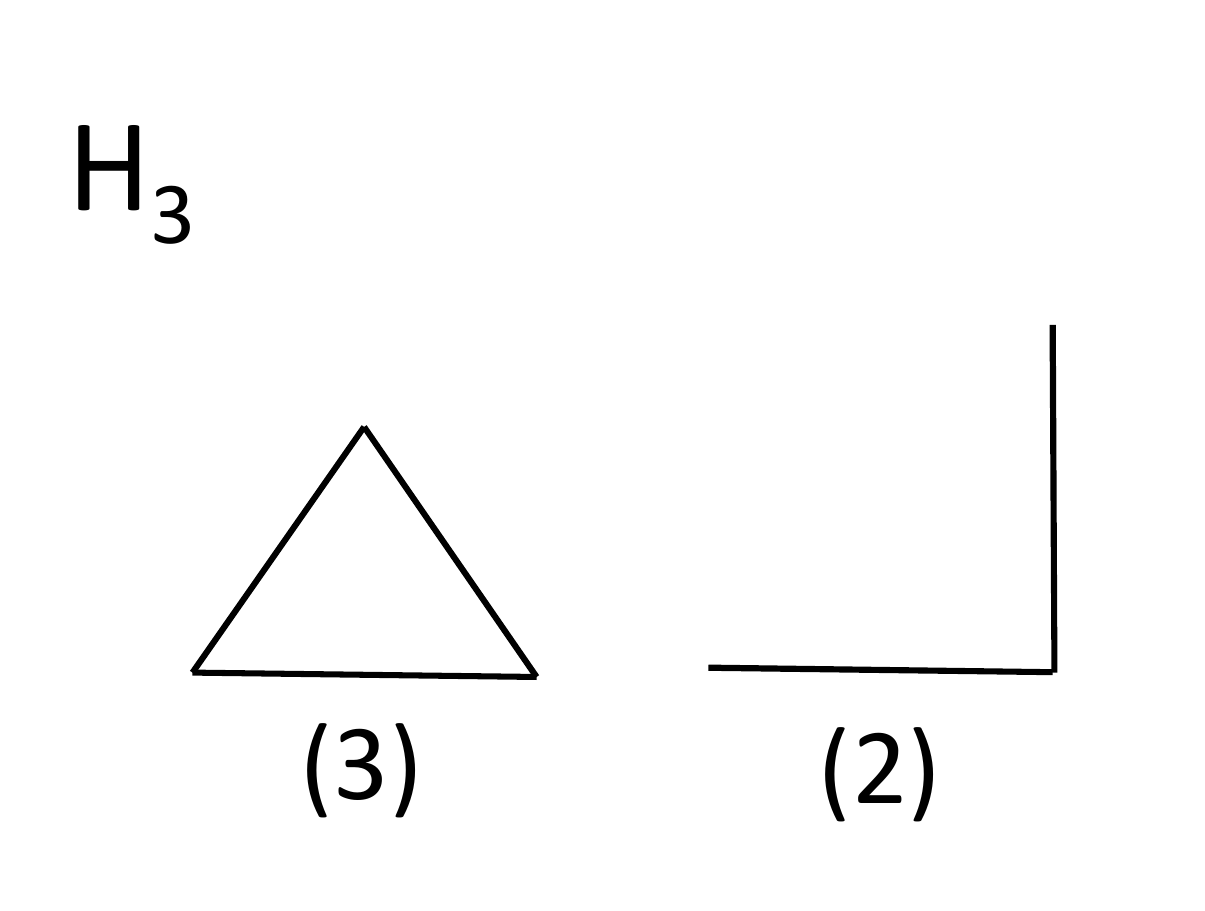}}
\caption{
The topological configurations of hypothetical $H_3$ molecules,
according to the number of links in a given configuration, (3) or (2).
These configurations can be formed in
$N_i =$ 1, 3 ways respectively. 
}
\label{f3}
\end{center}
\end{figure}           

If we consider a (hypothetical)  $H_4$-molecule, then the maximum number 
of connections is 6 and the minimum number (keeping still a bound cluster)
is 3. A molecule with 5 bounds can be obtained by cutting one of the
6 bounds, and this can be done 6 ways.  A molecule with 4 bounds can be 
obtained by cutting one more bound, as the lines are indistinguishable we 
do not take the order of which we remove the lines into account. 
This can then be done in 6*5/2, so altogether
this gives 15 configurations, but these will have two different topological
configurations:\\
(4A) 1 way with each node having 2 links, i.e. in tot. 
$N_{(4A)}=3*1$ ways and\\ 
(4B) 4 ways with nodes having 1, 2, 2 \& 3 links, i.e. in tot. 
$N_{(4B)}=3*4$ ways.\\
Then with a further cut (4A) leads to a linear chain with
3 links (3A) which can be obtained 4 ways,  i.e. in tot. 3*1*4/3 ways. The (4B) configuration leads to
the (3A) chain in 2 ways by 
cutting one of the links at the 3-link node, in tot. 3*4*2/3 ways. 
Thus the (3A) configuration can be
reached in total  $N_{(3A)}=3*4*3/3$ ways.\\
There is a further 3 link configuration which can be generated from
(4B) by cutting the link between the two 2-link nodes, this can only
be done one way, i.e. $N_{(3B)}=3*4*1/3$ ways in total.

\begin{figure}[h]     
\begin{center}
\resizebox{0.5\columnwidth}{!}
{\includegraphics{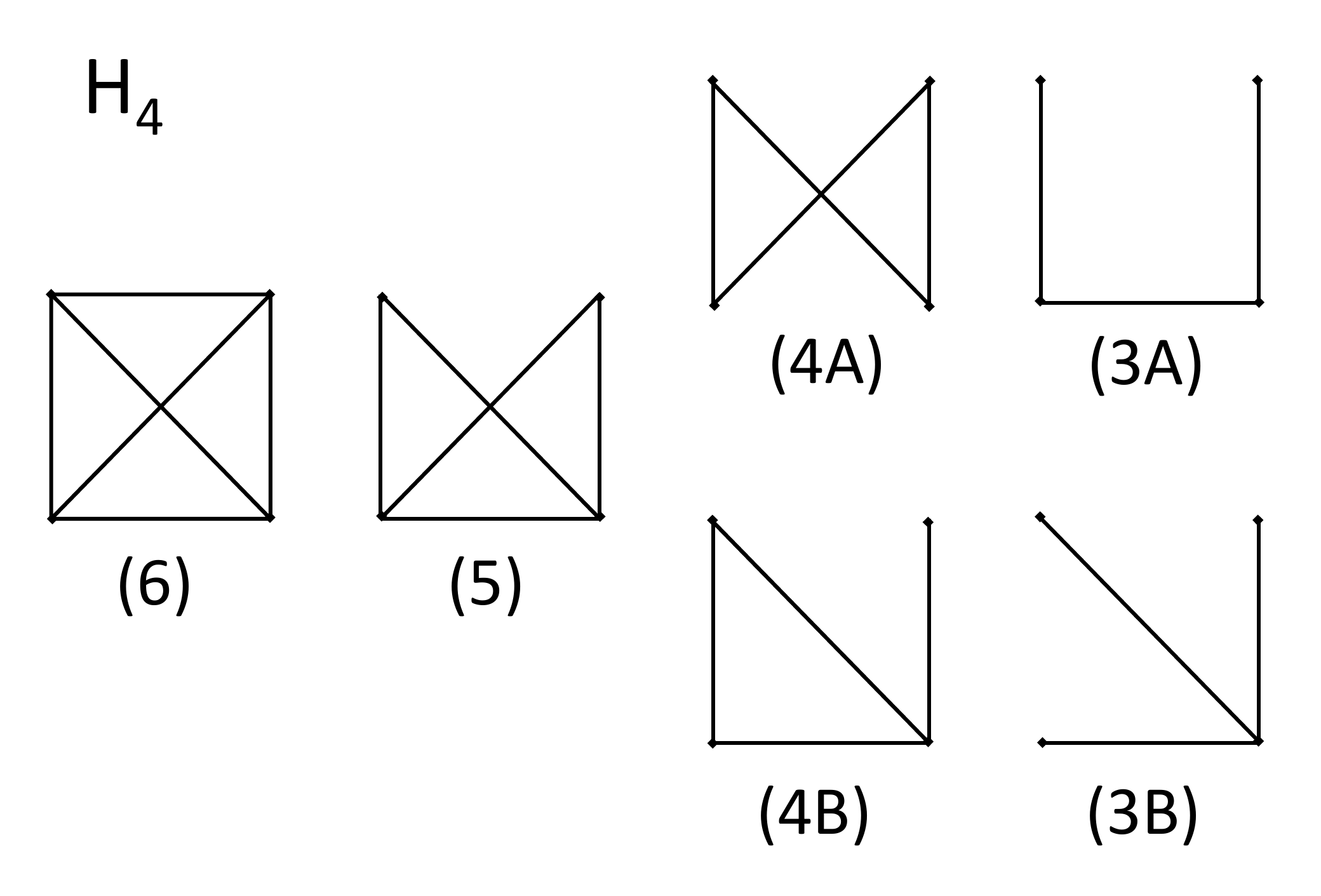}}
\caption{
The topological configurations of hypothetical $H_4$ molecules,
according to the number of links in a given configuration, (6), (5),
(4A), (4B), (3A), (3B). These configurations can be formed in
$N_i =$ 1, 6, 3, 12, 12, 4 ways respectively. 
}
\label{f4}
\end{center}
\end{figure}           

Thus the (6) and (5) link structures have only one type of topological
configuration, although with different topological formation probabilities,
the (3) and (4) link configurations have two types of topological 
configurations,
A and B, and these have different topological formation probabilities.
The sum of all configurations is $N = \sum_i N_i = 38$ and so, 
$p_i = N_i/N = 0.0263, 0.1579, 0.0789, 0.3158, 0.3158, 0.1053$.

If all these configurations are realized, with the above probabilities 
then the Shannon entropy of the system is
\be
H(X) = - \sum_{i=1}^N  p_i \ln p_i = 
- \sum_{i=1}^6  \frac{N_i}{N} \ln \frac{N_i}{N} = 1.5526 \ .
\ee
This entropy is a sum of all six possible configurations. The entropy 
of these six subsets is additive so we can also calculate the
entropy of a given configuration or of a subset of configurations.
Thus for the configuration X=(4A), we obtain
\be
H(4A) = 0.3662 \ .
\ee

Thus the connection or network topology should also be taken into account.
We can consider this on the hypothetical example of 
$H_1$,  $H_2$,  $H_3$, $H_4$ gases, with and without taking into account 
the topological entropy.

In real physical situations, not all (hypothetical) configurations are 
realized. Furthermore,
the links also have energies, so the experimental probabilities are not
exactly the same as the topological estimates. Nevertheless, this topological
example is able to give guidance on how to take topological 
(or network) structures into account in entropy estimates.

In the case of usual (Shannon) entropy estimates, the normalization is not the 
same as the physical one, but it is perfectly sufficient for comparative
studies of these types of structural entropies. 

Now the question arises, how do we add this configuration entropy
to the expression in eq. (\ref{eq1})? For this purpose we convert
both definitions into dimensionless form. The configuration
entropy, i.e. the Shannon entropy is already in
dimensionless form. Boltzmann's original definition for any
(non-equilibrium) system has also the same structure.
The definition for one particle
\be
H(X) = \sigma_P = \frac{S}{k_B\, N_P} = - \sum_{i=1}^N  p_i \ln p_i \ ,
\label{plnp}
\ee
may define both the sum for the probabilities of all configurations, $p_i^c$,
or the sum for all probabilities to be 
in a phase space volume element, $p_i^p$,
of size $(2\pi \hbar)^3$. This in this latter case is
\be
p_i^p = (2\pi \hbar)^3 f(x,p) \ ,
\ee 
where $f(x,p)$ is the phase space distribution
function normalized for one particle. In this latter case
$\sigma_P$ is the dimensionless specific entropy for
one particle (or molecule), which can be obtained from the
physical entropy $S$ as given by eq. (\ref{plnp}).

For a relativistic gas, 
the entropy density, using Boltzmann's definition for
any equilibrium or non-equilibrium phase space distribution, $f(x,p)$,  is
\cite{Csernai-book}
\be
s(x) = - \int\!\! \frac{d^3p}{p^0} p^\mu u_\mu \
f(x,p) \left[ \ln\left((2\pi \hbar)^3 f(x,p)\right){-}1\right],
\ee
where $p^\mu u_\mu$ is the frame invariant relativistic expression
for the local energy density. The last term, -1, ensures the 
appropriate entropy constant for joining smoothly the low temperature
quantum statistical limit in case of Boltzmann 
statistics\footnote{For Fermi-Dirac or Bose-Einstein distributions 
the additional term, -1, does not appear, but at the same time the calculation 
of the probabilities of the phase space cell occupations is more involved.}.

If we reach an equilibrium, we have a stationary solution of the Boltzmann
Transport Equation, e.g. the J\"uttner distribution,
 $f(x,p)$, and the entropy becomes 
\be
s(x) = - \int d^3p\ f(x,p)\ 
\left[\frac{\mu}{T}-\frac{\varepsilon}{T}-1\right]
\ .
\ee
Here $\mu$ is the chemical potential,
$T$ is the temperature and $\varepsilon$ is the 
specific energy, the energy for one particle.
Notice that in the relativistic theory both $\mu$ and $\varepsilon$
includes the rest mass of the particle, but in the entropy expression
these terms cancel each other. In the non-relativistic limit for the
Boltzmann distribution this leads to the entropy expression of
eq. (\ref{eq1}).

If the phase space distribution is normalized to $n_P(x)$, i.e.
\be
n_P(x) = \int\! d^3p\ f(x,p) \ ,
\ee
then the entropy for one particle can be obtained for the Boltzmann 
statistics as
\be
\sigma_P^{ph.s.} = \frac{s}{k_B\, n_P} = 
- \sum_{i\epsilon\ ph.s.}  p_i^p [ \ln p_i^p - 1 ]
\ee
where $p_i^p = (2\pi \hbar)^3 f(x,p)$, the probability to be in a phase
space cell, $i$, should be calculated for one
particle, i.e. 
\be
\int\! d^3x\ n_P(x) = 1 \ .
\ee
Still the entropy, the distribution function and $p_i^p$ depend
on the particle density, $n_P$.

Thus in conclusion the single particle entropies should be additive
in a configuration
\smallskip
\be
\sigma_P =  \phantom{mm_I}
\sigma_P^{conf.} + \sigma_P^{ph.s.}  
\phantom{mmmm_I}
=
-\!\!\!\! \sum_{i\epsilon\ conf.}\!  p_i^c \ln p_i^c  
-\!\!\!\! \sum_{i\epsilon\ ph.s.}\!  p_i^p [ \ln p_i^p - 1],
\label{ee14}
\ee
where $p_i^c$, is the probability to have a configuration state, $i$.
In conclusion the increase of the degrees of freedom due to the
different configurations leads to an increase in the single particle entropy.
This increase is very small if the number of possible configurations,
$N = \sum_i N_i$ is large, but the number of a realized configurations,
$N_i$, is much smaller, $N_i \ll N$.

\begin{table}[h]   
\begin{center}
\begin{tabular}{crrrr} \hline\hline \phantom{\Large $^|_|$}
	Material\ \  &  $A_P$\ \  & $N_P$\ \ &  $\sigma_P$\ \  & $S_{1kg}$\ \ \\
	      &    & (mol)  & &  (J/K)  \\
\hline
$H_1$       & 1 &\ \ 992.092 &\ \ 13.106 &\ \ 108111.7 \\
\smallskip\smallskip
$H_2$       & 2 & 496.046 & 14.146 &  58344.0 \\
$H_3$       & 3 & 330.697 & 14.754 &  40568.3 \\
$H_3^{(3)}$ & 3 & 330.697 & 15.101 &  41521.2 \\
\smallskip\smallskip
$H_3^{(2)}$ & 3 & 330.697 & 14.970 &  41161.6\\
$H_4$       & 4 & 248.023 & 15.186 &  31316.1 \\
$H_4^{(6)}$ & 4 & 248.023 & 15.282 &  31513.5 \\
$H_4^{(5)}$ & 4 & 248.023 & 15.477 &  31917.1 \\
$H_4^{(4A)}$& 4 & 248.023 & 15.386 &  31729.5 \\
$H_4^{(4B)}$& 4 & 248.023 & 15.550 &  32066.8 \\
$H_4^{(3A)}$& 4 & 248.023 & 15.550 &  32066.8 \\
$H_4^{(3B)}$& 4 & 248.023 & 15.423 &  31804.8 \\
\hline
\end{tabular}
\end{center}
\caption{
Entropies of a single composite particle and of 1kg material 
in different topological configurations for hypothetical
$H_1$, $H_2$, $H_3$ and $H_4$, molecules 
approximated as ideal gases, depending on the mass
numbers, $A_P$, of the nucleons in the molecule and the
configuration where it is indicated. 1kg material contains 
different mol-numbers of particles, $N_P$, and this number is
decreasing with increasing $A_P$. The specific entropy
of the composite particles is indicated by the dimensionless
($\hbar,c,k_B=1$) $\sigma_P$. The total entropy of
the 1kg material at STP is also decreasing with increasing
complexity, i.e. increasing $A_P$.
}
\label{t2}
\end{table}   

In the above discussed estimate we only considered differences in
topological configuration among identical particles. In real situations
the complexity may increase due to: direction dependence of the links,
different constituents, different (energetic) weights
of the links, dynamical freedom of the  length or angle of the link, 
etc.

In Table \ref{t2} the hypothetical ideal gas particles, $H_1$ and $H_2$,
have no option for different
configurations. For $H_3$, the specific entropy, $\sigma_P$, increased by
a relatively small amount of 0.347 and 0.216 for the topological
configurations $H_3^{(3)}$ and $H_3^{(2)}$ respectively. The entropy of the
different $H_4$ configurations is increased by small values of configuration
entropy between 0.096 and 0.364. The increase in configuration entropy 
is very small for configurations with small probability, $p_i$, which
may occur for complex systems with a large number of possible configurations
where only a few are realized in a realized sample of configurations.

The configuration entropy may be much larger for more involved structures,
where the number of configurations are comparable or may even exceed the
number of particles. Also, an important question to ask is how many
of these configurations can be realized, and how many are actually 
present in a given sample we discuss.

\section{Entropy of Phases of Physical Systems}
\label{S3}

Physical systems may have a variety of configurations, different
(i.e. nonidentical) constituents, and different 
physical degrees of freedom, such as vibration, rotation, etc.,
in addition to the phase space occupancy. Because of this,
their entropy may well exceed the entropy based on the ideal
gas approximation. E.g. 1 kg ideal gas with mass number 18 ($H_2O$)
has an entropy of $S_{1kg} = 8243.0$ J/K, at the boiling point, while the
physical value is about 10495 J/K. This can be explained by the
different types of constituents, $H$ and $O$, the configuration 
and other dynamical degrees of freedom and types of interactions.

\begin{figure}[h]      
\begin{center}
\resizebox{0.6\columnwidth}{!}
{\includegraphics{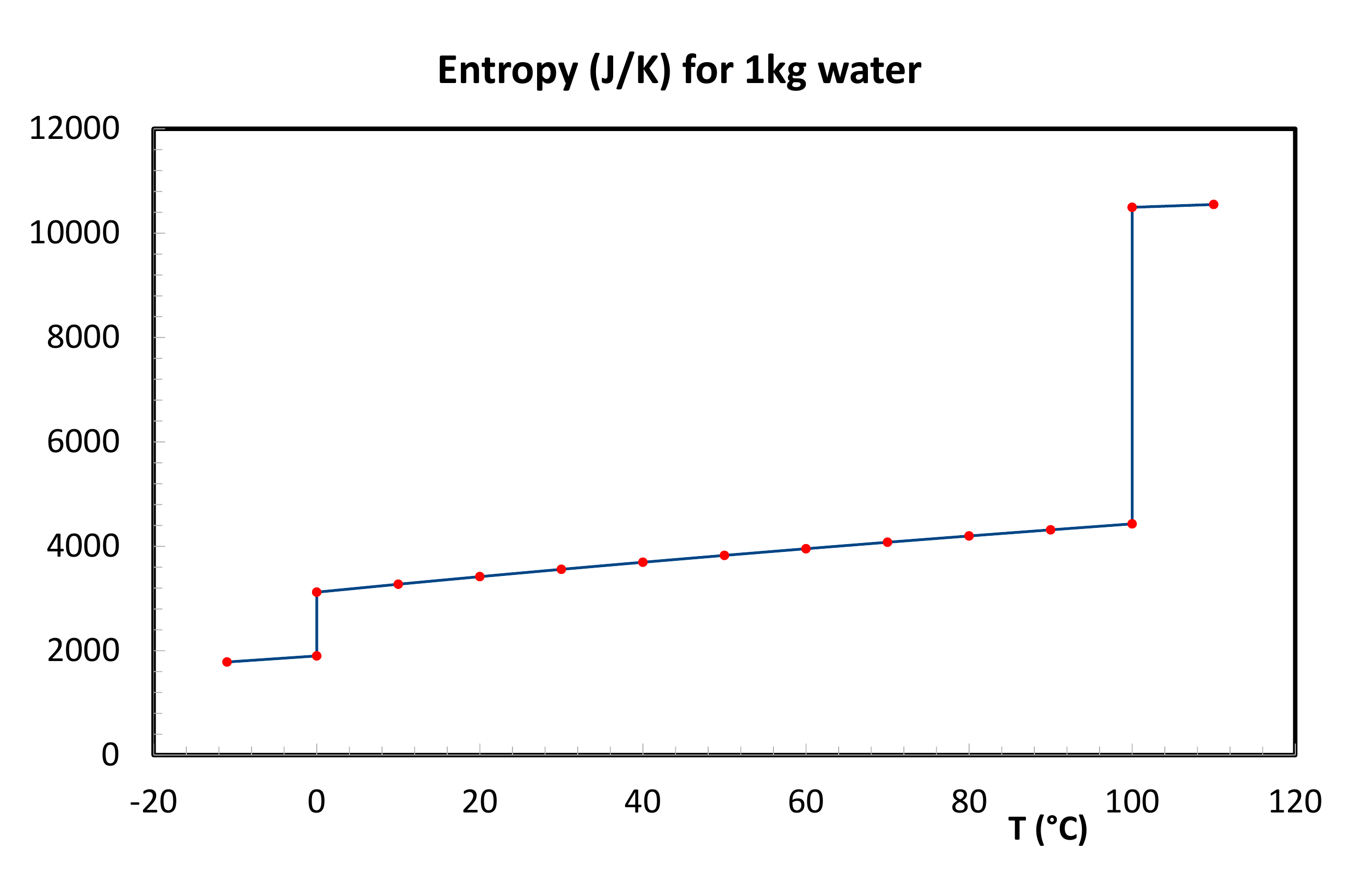}}
\caption{
(color online)
Entropy of 1 kg of water at different temperatures, with all three
phases, ice, water and vapour, with two phase transitions at
0 and 100$^oC$. The reference point is taken at $T=100^oC$ liquid
water, with entropy of $S_{1kg} = 4430.01$ J/K, where the 
phase transition to vapour with a latent heat of $\Delta S = 6065.55$ J/K,
leads to a vapour entropy of $S_{1kg} = 10495.56$ J/K.
Using the water heat capacities we reach the phase transition between
water and ice with water entropy of  $S_{1kg} = 3122.92$ J/K and
ice entropy of $S_{1kg} = 1900.15$ J/K.
}
\label{F3-RY-vs-t}
\end{center}
\end{figure}        

\begin{figure}[h]     
\begin{center}
\resizebox{0.4\columnwidth}{!}
{\includegraphics{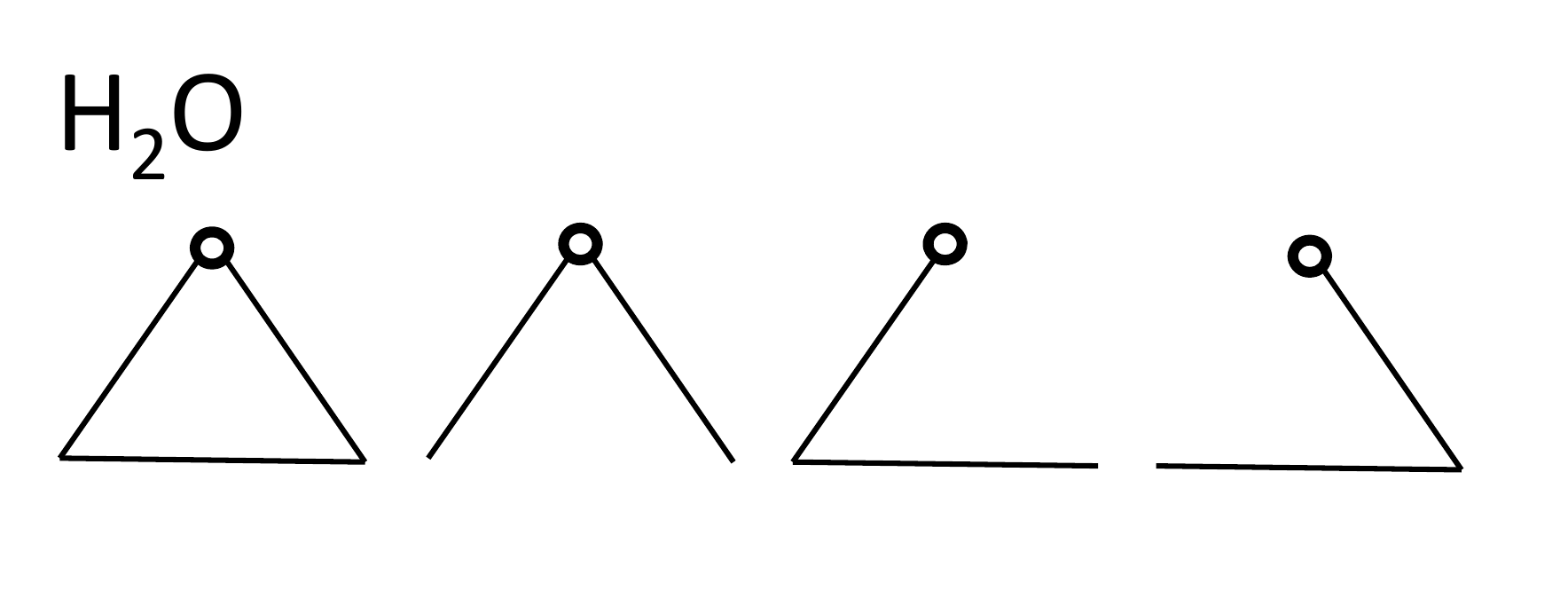}}
\caption{
The topological configurations of a $H_2O$ molecule.
The circle indicates the Oxygen atom.
If the two $H$ atoms are identical then the last two
configurations are identical also.
}
\label{f5}
\end{center}
\end{figure}

If we add the topological configuration entropy,  $\sigma_P^{conf.}$,
to the ideal gas estimate of the $H_2O^*$ molecule (at $T=100^o$C), 
we get a larger value. 
Here we have two possibilities, the two Hydrogen atoms
can be identical (or not). In these two cases we have 3 (or 4) possible 
topological configurations, see Fig. \ref{f5}.
The specific topological configuration entropies,
$\sigma_P^{conf.} $, can be calculated as 
in section \ref{S2}. Here we assume equal probability for each 
topological configuration.
There are four possible cases: either all
possible configurations are realized or only a single one, and either
the $H$ atoms are identical ($N=4$ possible cases) or not 
($N=3$ possible cases). Then, the entropy can be calculated as:

\be
\sigma_P^{conf.} = - \sum_{i=1}^N  p_i \ln p_i = \ln N
\ \ \ \ \ \ \ {\rm or} \ \ \ \ \ \ \ \ 
\sigma_P^{conf.} = - p_i \ln p_i ,
\ee
for all configurations or a single one respectively. These
additive topological configuration entropies are given in Table \ref{t3},
together with the total entropy values where the entropy
from the phase space occupation, $\sigma_P^{ph.s.} = 17.988$ and 
for the  topological configuration are added up. 
\begin{table}[h]   
\begin{center}
\begin{tabular}{crrrr} \hline\hline \phantom{\Large $^|_|$}
	\ \  &  All 4  & All 3 &  One of 4  & One of 3 \\
\hline
$\sigma_P^{conf.}$\phantom{\Large $^|$}
              &  1.386 & 1.099  &  0.347 & {\bf  0.366} \\
$\sigma_P$    &  19.374 & 19.087 & 18.335 & {\bf 18.354} \\
\hline
$S_{1kg}^{conf.}$ (J/K)\phantom{\Large $^|$}
	               & 635.1 & 503.63 & 159.02 & {\bf 167.7}\\
$S_{1kg}$   (J/K)      & 8878.3 & 8746.8 & 8402.2 &  {\bf 8410.9} \\
\hline
\end{tabular}
\end{center}
\caption{
Topological configuration entropy and total entropy of $100^o$C 
water vapor. The realistic physical case is in the last column. 
}
\label{t3}
\end{table}   

If all three (or four) configurations would be realized with the
same probability then we get a larger final specific entropy,
$\sigma_P = \sigma_P^{conf.} + \sigma_P^{ph.s.} = 19.087 (19.374)$ 
and the corresponding entropy for 1 kg material will become
$S_{1kg} = 8746.6\ (8878.3)$ J/K.  
If only one configuration is realized
from the three (or four) possible ones, then the final entropies are smaller.
For the specific entropy 
$\sigma_P = 18.354\ (18.335)$
while for 1 kg of material
$S_{1kg} = 8410.9\ (8402.1)$ J/K. These last smaller values indicate
that the larger information content (i.e. that only one state is
realizable from the possible ones) decreases the entropy of the 
given material.

We see that the estimate that includes the addition of topological entropy
moved the estimated entropy of the system closer to the experimentally
observed value. Obviously, we did not reach the experimental value
as several other degrees of freedom (angles, lengths, etc.) were
not taken into account, nor the energetic configuration of the
water molecule in the vapour.

The exact absolute values of ice are less well known, because
the ice has several different configuration structures and it
is difficult to make measurements down to $0^o$K temperatures.
The theoretical calculation of complex energetic structures is
difficult and the theoretical calculations of structural changes
of a phase transition is even more so.

At the phase transitions the heat capacity has a pole, which is
again problematic to calculate, so the exact "calculation" of absolute
entropies is already difficult for relatively simple molecules 
with a phase transition.

Nevertheless, we attempt to estimate the absolute entropy values
of more complex molecules also, where experimental entropy 
measurements do not exist. {\it There are no fundamental reasons
to prevent us from estimating the absolute entropies of highly 
complex molecules.}

For the fundamental determination 
of absolute entropy values it is important that (i)
we quantized the phase space cells with a well defined physical
value, and  (ii) we connected the entropy 
to a well defined amount of matter and its energy.

The role of configuration entropy is in principle clear.
The 3 atoms of water belong to the above discussed configurations,
but the weight of the different probabilities for these configurations
are of course different because the connections carry energy, which
depends on the "angle" of the two H atoms in the $H_2O$ molecule.
Or more precisely, these depend on the characteristic angle of the 
quantum mechanical
wave function of the molecule. In different model calculations
this angle varies between $104.52^o$ and  $109.5^o$, and it may also 
be dependent on the density and temperature of the material.

This feature indicates that the probabilities of the real physical
configurations may not be the same as in the oversimplified 
configuration estimates presented in the previous section.  
Some configurations can have a highly suppressed probability, or
might not be realized at all in nature. E.g. for water the probability
of first configuration in Fig. \ref{f5}, is probably much smaller 
(based on the wave-function) than in the above topological estimate.

We will see that in more complex structures, there are possible 
configurations which do not appear in nature at all. To 
find out which configurations are realized and with what probability,
is a much more difficult question than just counting up the 
number of possible configurations.
\bigskip

{\bf The ratio of configuration entropy to kinetic entropy:}\ \
In Table \ref{t1} one we could see that the kinetic entropy increases with
increasing particle mass, but this increase is rather slow. Furthermore,
this table was calculated with the ideal gas approximation i.e. like the water vapour
phase of water. If we consider liquid water at a lower temperature, e.g. room temperature,
the entropy of the system is reduced by about 60\%, 
as shown in Fig. \ref{f5}.
While for water vapour the ratio of kinetic entropy to configuration 
entropy is roughly 20:1, for liquid water at room temperature it
is only about 5:1~! With more complex molecules the weight of configuration 
entropy increases further and it may overcome the kinetic entropy.

\section{Entropy of Complex Molecules}
\label{S4}

As the example of water indicates, more complex molecular structures
decrease the entropy compared to the ideal gas approximation. This is due to 
the interaction among the atoms of a molecule, as well as the interactions
among the molecules in a given amount of material. Both these effects
decrease the available phase space (e.g. compared to the ideal gas), and
therefore these interactions also decrease the entropy. Especially the 
momentum space degrees of freedom are reduced as the interactions hinder
the possibility of large momentum differences among the atoms of a molecule
as well as among the molecules.

Furthermore, the atoms of a complex molecule may take different spatial
configurations. This was not considered in the previous topological 
configuration estimates. The spatial configurations of the molecules 
depend also on the surroundings, so that a given molecule may have
different configurations with different entropies. Especially in biology,
these different configuration possibilities have important physiological
roles. In case of molecules, these configurations are determined energetically,
and can (in principle) be calculated based on the interaction energies
within the molecule and among the molecules. Thus, the quantization of the
phase space still enables us to obtain the absolute entropy of a given
configuration. At a given temperature, one can also determine which 
configurations are realized, and with what probability. An example of such
calculations is the calculation of protein and other molecule conformations, 
which is a widespread activity today.

The entropy of such a complex molecule is estimated indirectly via the
Helmholtz free energy, $F$, where with the inverse temperature, 
$\beta = 1/(k_B T)$ we can calculate
\be
\beta F(T,V) = - S + E/(k_B T) .
\ee
The effect of interaction is estimated via the correlation function
of the density \cite{LB90}, by considering the atoms as hard spheres.

The same type of approximation is used in ref.
\cite{MSSWF16}, where the possible stable (or energy minimizing) packings
were counted, and the Gibbs and Edwards (Boltzmann) entropy values were
evaluated as a function of the system size.
Obviously the number of possible configurations is much larger this
way than in the previous topological estimates, but at the same time
the locally energy minimizing configurations are considerably fewer
than all configurations.

By counting the disordered 3D sphere packings this way the configuration
entropy in the spatial occupation can be estimated. 
%
In the context of granular packings, our
aim is to compute the number of ways, $\Omega$, in which 
$N$ spheres can be arranged in a given abstract volume of 
dimension $d$.  Then 
the total available abstract volume in $dN$-dimensional
space is  $\mathlarger{\mathlarger{\nu}}$.
%
We can consider the
volume of the basis of attraction of each atom in the molecule, 
in a distinct energy minimum 
\be 
\mathlarger{\mathlarger{\nu}} = \sum_{i=1}^\Omega \nu_i ,
\ee
where $\nu_i$ is the volume of the $i$-th basin of attraction,
$\Omega$ is the total number of distinct minima and 
$\mathlarger{\nu}$ is the accessible volume.
Thus,
\be
\Omega = \frac{\mathlarger{\mathlarger{\nu}}}{\langle \nu \rangle} ,
\ee
where $\langle \nu \rangle$ is the mean basin volume.
Then the (Gibbs) entropy can be obtained as
\be
S_G = - \sum_{i=1}^\Omega p_i \ln p_i  - \ln(N!) \ ,
\label{eq19}
\ee
where $p_i = \nu_i / \mathlarger{\mathlarger{\nu}}$.
A basic estimate is that the
dimensionless entropy is 
\be
S_G \approx N/2   \ ,
\label{eq20}
\ee
for molecules of up to N=100  atoms
\cite{MSSWF16}. Then the entropy increases slower.
So, we estimate that for very large molecules
the specific spatial configuration entropy
increases slower, e.g. as $ \propto \ln N$.


In case of a given bio-molecule, like the DNA, not 
all spatial configurations are realized.  Consequently the
sum in eq. (\ref{eq19}) is reduced to a single configuration or to
a few configurations, which results in a specific entropy
of the order 
\be
S_G = -  p_i \ln p_i  - \ln(N!)  \approx
\frac{\langle \nu \rangle}{\mathlarger{\mathlarger{\nu}}} 
\ln \left(
\frac{\langle \nu \rangle}{\mathlarger{\mathlarger{\nu}}} 
\right) \  ,
\label{eq21}
\ee
and since $\mathlarger{\mathlarger{\nu}} = N \cdot \langle \nu \rangle$
\be
\sigma_G  \approx
\frac{1}{N} 
\ln \left(N\right) \  .
\label{eq22}  
\ee

Therefore, in the specific entropy per 1kg or per 
atomic number, the contribution is of the order of one
and decreasing with increasing number of atoms, $N=N_a$, in the molecule.

As in most of the literature, these approaches give an entropy value
which serves well for the comparison of different complex molecular 
configurations. At the same time, an absolute entropy value for a given
complex molecule and for a given spatial configuration is not given,
although this could in principle be possible.

%
The physical entropy of a complex molecule can be calculated 
based on all degrees of freedom in the phase space, and  
all interactions among the constituent atoms. In real situations,
to calculate all possible realized and not realized configurations 
with their energies, is beyond our possibilities. So many different
approaches were constructed based on the interconnected network
of the constituents \cite{Cserm12}. This way static entropies
may characterize the global disorder of network
topology. One of the approaches to define network
entropy is based on the number of the neighbors of a node.
This is a degree-based entropy, meaning it is low when the degree
of the nodes is uniform and small. 
There exists more complex definitions of entropy, which account
for the interaction between all possible pairs
of nodes.  The analysis of the entropy of representative 
subnetworks has also become possible.

Large molecules can form {\bf polymers}, or even larger 
macromolecules, which are composed of many repeated subunits. 
These may have a broad range of properties in the case of both synthetic 
and natural polymers. In these cases the previous dense packing 
approach is not applicable.

\section{DNA in Bacteria}
\label{S4b}
The level of complexity is increased in 
live structures that can replicate themselves. 
The basis of this replication is the 
deoxyribonucleic acid (DNA) molecule,
which structure is a double spiral.
The DNA stores a code made of four chemical bases: 
adenine (A), guanine (G), cytosine (C) and thymine (T).
The sequence of these bases determines the information available 
for building and maintaining an organism. DNA bases pair up with 
each other, A with T and C with G, to form units called base pairs.
Each base is also attached to other molecular structures 
which together, the base and its support, is called a nucleotide.
The DNA can replicate, or make copies of itself. Bacteria, 
as asexual organisms, inherit identical copies of their 
parent's genes.

The average weight of a DNA base pair (bp) is 650 daltons or 650 AMU
(where 1 AMU 
= $1.660539040(20)\cdot 10^{-27}$ kg 
= $931.4940954(57)$ MeV/c$^2$.) 

One of the smallest live systems is the 
endosymbiotic 
\footnote{Endosymbiont is an organism that lives within the 
body or cells of another organism.}
bacteria Candidatus Carsonella ruddii (CCr), which has
$N = 159 662$ base pairs.  Its genome is built up by a circular chromosome.

The possible number of configurations of this number of 
{\bf base pair sequences} 
is $4^N$, so the probability of the single existing CCr sequence is
$p_i = 4^{-N}$. The corresponding 
base pair sequence entropy is then
\be
\sigma_P^{bp}\ \  =\
H(X)\ =\ -p_i \ln p_i = -4^{-N} \ln 4^{-N} \  =\  N 4^{-N} \ln{4}\ =\ 1.386\, N \,4^{-N}\ ,
\label{q23}        
\ee
thus 
\be
\ln H(x)\ \ =\ - (1.386 N) + \ln{N} + \ln{1.386}
         \ =\ - 221291.532 + 11.981 + 0.326 
         \ =\  - 221279.225 \approx\ \  - 2.2\cdot 10^5 \ .
\label{q24}
\ee
Thus, the specific entropy of the CCr DNA molecule on the base pair sequence
configuration is
\be
\sigma^{bp}_P \approx e^{-221\,279} = 10^{-96\,099} \ .
\ee

The DNA of the CCr is 159\,662 base-pairs and its weight is 
$ m_{DNA_{CCr}} = 159\,662 \cdot 650$ AMU $ = 7.727\cdot 10^{-18}$ kg.
So, the number of CCr DNA molecules in 1kg is  
$N_{DNA_{CCr}} = 1/m_{DNA_{CCr}} = 1.294\cdot 10^{17}$.

The entropy of 1 kg of the DNA molecules based on their b.p. configuration
 is then 
\be
S_{1kg-CCr}^{bp} \ \ = 
k_B\, N_{DNA_{CCr}}\, \sigma_P^{bp} =\ \  1.786\cdot 10^{-96\,105} {\rm J/K} \ .
\label{Sbp}
\ee

\smallskip

This DNA in the CCr bacteria builds up circular chromosomes. In the physical phase space
the momentum part is largely negligible but the {\bf spatial configuration} 
is substantial. For an exactly given sequence of length, in this case 159 662 base pairs,
the number of possible spatial configurations is large. It builds a 
circular chromosome, but even if the sequence is fixed, the shape can be
different. For cell replication one might need a more straight 
configuration, and other more compact configurations are also possible
although these might be energetically less favourable. Here we cannot 
estimate energetically all possible configurations and their probability.
Instead, we make an overly simplified estimate.

The stretched out DNA could expand a sphere of radius $R=N/2$
with a volume of $V = 4 \pi R^3 / 3 =  V = \pi N^3 / 6 = 8.673\cdot 10^{30}$.
If we position the 1st base-pair to a certain location, the next base pair can be put to
$\sim$ 25 neighbouring points (if these are not occupied). As the volume is
25 orders of magnitude larger than the number of base-pairs, we can
neglect the possibility that a chosen point is already occupied.
Thus the estimated number of all connected spatial configurations can be 
\be
M = 25^{N}
\ee
Here we assumed that the base-pairs can be connected in any angle
within a cubic grid.

The number of circular configurations is much smaller. The circle has a 
circumference of N and the diameter of the circle is $D = N/\pi$.
Let us choose a random point on the circle. Next, let us choose the
opposite point, which is on a sphere of $4 \pi D^2$. These two points 
give an axis of the circle. The third point should set the plane of the
circle, which is on a circle of length $ D \pi$.
After we have chosen these three points, the circle is fixed and there is no 
more freedom. So the number of possible circles is
\be
M_c \ \ = 4 \pi D^2  \times D \pi = 4 \pi^2 D^3 = \ \ 4 N^3 / \pi \ .
\ee
Thus the probability of an arbitrary spatial configuration is
$ p_i = 1/M$, and the 
spatial configuration (s.conf.) entropy is
\be
\sigma_P^{s.conf.} 
\ \ =\ H(X)\ =\ - \sum_{i=1}^{Mc} p_i \ln p_i\ =\ \frac{M_c}{M} \ln M
\ =\  \frac{4 N^3}{\pi 25^N} \ln 25^N\ =\ \frac{4 N^3}{\pi 25^N} N \ln 25
\ =\ \frac{4\, 3.219}{\pi} 25^{-N} N^4\ =\ \ 4.099  N^4 25^{-N}
\ee
thus
\be
\ln H(X) 
\ \ =\ 1.411 + 4 * \ln N  - N \ln 25 
\ =\  - 513902.644 \approx \ \ - 5.139 \cdot 10^5 \ .
\ee
Thus, the specific entropy of the CCr DNA molecule on the spatial 
configuration is
\be
\sigma^{s.conf.}_P\ \  \approx\ e^{-513\,903} =\ \ 10^{-223\,183}
\ee

This is a similarly low specific entropy as the one from
the base-pair sequence, eqs. (\ref{q23}-\ref{q24}).  The entropies 
from these two independent degrees of freedom
should be added
\be
\sigma_P\ \  = \sigma_P^{bp} + \sigma_P^{s.conf.} \ ,
\ee
where actually the larger $\sigma_P^{bp}$ dominates,
so that 
\be
\sigma_P^{DNA}\  \approx\ \ 10^{-96\,099}\ ,
\ee

The entropy of the DNA molecules in 1 kg  matter is then 
\be
S_{1kg-CCr}^{DNA}\ \  = \
k_B\, N_{DNA_{CCr}}\, \sigma_P^{DNA}\ = 
\ \ 1.786\cdot 10^{-96\,105} {\rm J/K} \ .
\label{SDNA}
\ee
This is the same as the entropy of the DNA b.p. configuration only,
eq. (\ref{Sbp}), because the entropy
from the spatial configuration is utterly negligible.

If we would count the spatial configuration only that would give a smaller
specific entropy value, and the resulting entropy for the DNA molecules 
would yield a smaller value
\be
S_{1kg-CCr}^{s.conf.}\ \  = \
k_B\, N_{DN\!A_{CCr}}\, \sigma_P^{DN\!A} \ =\ \  1.786\cdot 10^{-223\,191} {\rm J/K} \ .
\ee

\section{Entropy of the CCr bacterium}

Without its {\bf environment} the DNA molecule cannot exist. 
If we want to calculate
the entropy of one kg material we have to estimate the other surrounding
constituents of in the CCr bacteria, in comparison with the DNA in the cell.

The total weight of a CCr bacterium can be estimated to 
$m_{CCR} \approx 2.3\cdot 10^{-17}$kg, i.e. about three times bigger than
the weight of the DNA, so two thirds of this weight is made up of smaller
molecules than the DNA. 
The number of CCr bacteria in 1 kg is 
$N_P = m_{CCR}^{-1} = 0.4\cdot 10^{17}$/kg.
Two thirds of this matter is approximated to be water, and one third 
of the weights is given by the DNA molecules.

The number, and thus the entropy, of the DNAs in 1 kg  CCr 
bacteria is then about three times less than estimated in eq. (\ref{SDNA})
\be
S_{1kg-CCr}^{DN\!A}\ \  = \ 
k_B\, N_P\, \sigma_P^{DN\!A}\  = \ \ 0.595\cdot 10^{-96\,105} {\rm J/K} \  .
\ee
\smallskip

For a first, extremely rough estimate, we can assume that this extra
material has about the same entropy as water, $S_{1kg}^{H_2O} \approx 4000 $J/K.
If we then calculate the entropy of 1 kg CCr bacteria:
\be
 S_{1kg}^{CCr} \ \ =\ \frac{2}{3} S_{1kg}^{H_2O} + S_{1kg}^{DN\!A}
 \  =\  2666\, {\rm J/K} + 0.595\cdot 10^{-96\,105} {\rm J/K} \  .
\ee
Thus, the contribution of the DNA would be utterly 
negligible to the total entropy
of 1kg CCr bacteria. The replacement of smaller molecules 
by water overestimates the entropy unrealistically. 

A  more realistic estimate is that the DNA leads to the build up of different
molecules in the cell. Their number and variation depends on the DNA,
but also on the environment (!). Due to the larger number and weight of 
these other molecules of large variation and complexity, the possible 
number of variations increases, compared to the possible variations
of the DNA structure. At the same time, the number of realized
configurations can be an even smaller proportion, due to the additional selection
caused by the environmental conditions. In conclusion, we 
can estimate that the complexity of all the smaller molecules of the cell
is larger, and their entropy is smaller, than the one estimated from the
complexity of the DNA molecule itself.

\begin{figure}[ht]     
\begin{center}
\resizebox{0.7\columnwidth}{!}
{\includegraphics{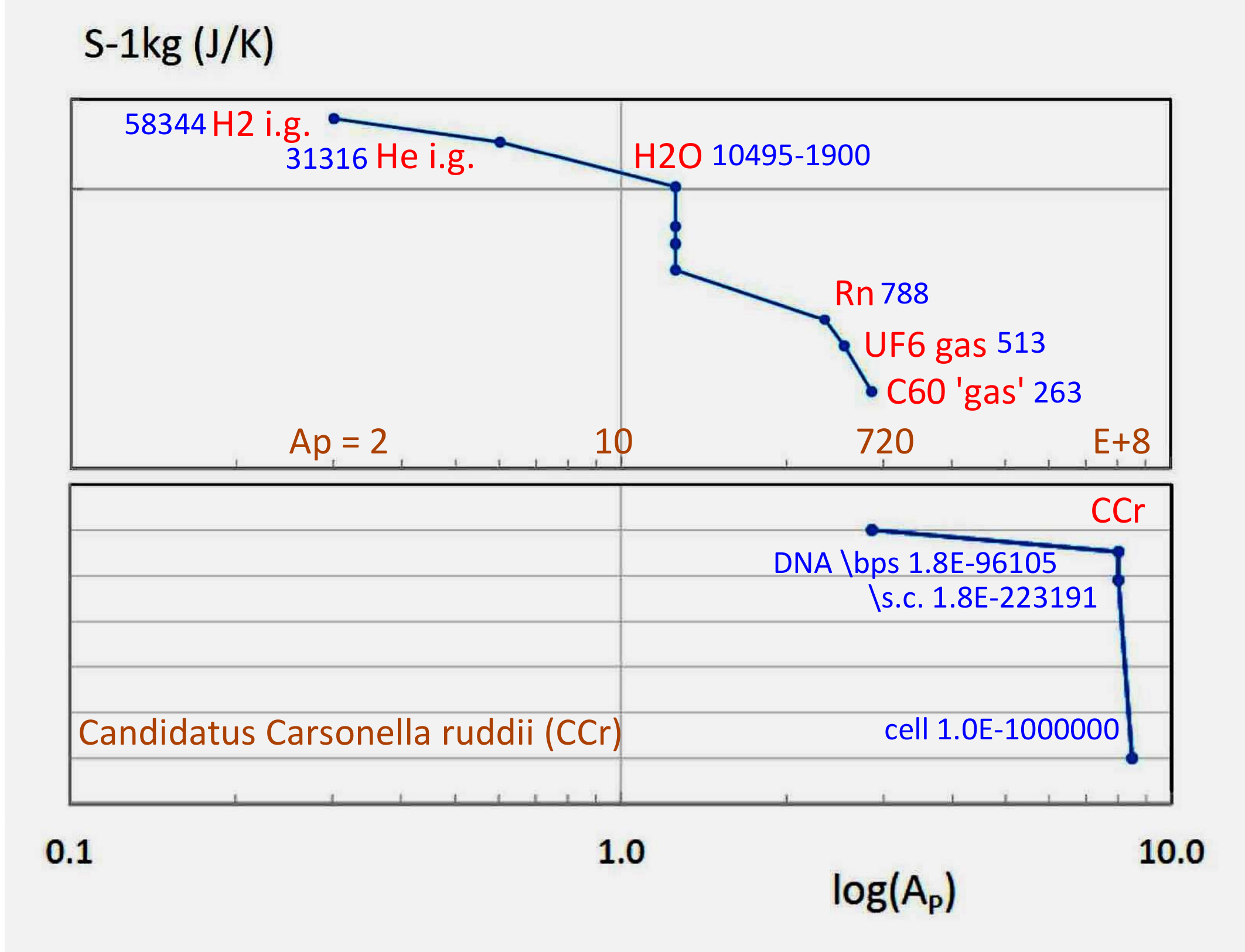}}
\caption{
(color online)
The entropy of 1kg material, $S_{1kg}$ in units of (J/K) from light 
molecules behaving like ideal gases up to the DNA of the simplest
bacterium, Candidatus Carsonella ruddii.  
This last one is based on its base-pair sequence (bps), its spatial 
configuration (s.c.), and all other different molecules inside the cell. 
The entropy values are plotted versus the atomic number of the molecules, 
which is the weight of the molecule in daltons. 
}
\label{f7}
\end{center}
\end{figure}        

Based on this reasoning, for unit mass,
the entropy of all the smaller molecules may be 
even smaller than the entropy of the DNA, so the more realistic estimate
for the entropy of 1 kg CCr bacteria is
\be
 S_{1kg}^{CCr} \ \ = \ k_B\, N_P\, \sigma_P^{DNA} \
 =\ \ 1.0\cdot 10^{-1\,000\,000} {\rm J/K} \  .
\ee

It is still amazing to estimate and see, how extremely small the entropy
of the most complex molecules and structures is.

\section{Entropy of Highly Complex Molecules}
\label{S5}

The building blocks of DNA are the nucleotides,
Thymine, Adenine, Guanine and Cytosine.
Each Thymine connects to an Adenine and each Guanine connects to a Cytosine.
This is one {\it allele}. Such a pair is called a nucleotide {\it base pair}.

The number of DNA base pairs (DNA-bp) in the total human genome 
(in the 23 chromosomes) is approximately 3.2 billion 
($A_P^{DNA-bp} = 3.234 \cdot 10^9$).
Each chromosome is a curled up
DNA molecule. In these DNAs there are some 
20000-25000 genes that contain these base pairs. 
Genes are sections of the DNA which act as instructions 
to make molecules, e.g. proteins.
Genes with a smaller
amount of base pairs (smaller DNA section) will have a higher entropy 
per mass while a longer DNA section will decrease the entropy.

The human body contains approximately 10 trillion ($10^{13}$) cells and each 
of those typical cells contain  the genome, i.e. 23 chromosome pairs. There is 
then approximately $6.4 \cdot 10^{22}$ base pairs in total, but since the 
genome is the same in each cell, the cell count does not increase
the sequential configuration entropy from DNA. 

The total amount of all sequences is $N_{all-sequences} = 4^N $,
where $N$ is the number of base pairs, 
$$
N_{bp} = A_P^{DNA-bp} \ .
$$
Assuming that all sequences are of equal probability, 
then for a given sequence, $k$, the probability is  
$p_k = 1/4^{N_{bp}}$ and we get 
\be
H_{max}\ \  =\ \ln 4^{N_{bp}}\ =\ N_{bp} \ln(4)\ = \ \ 4.482\cdot 10^9
\ee
by adding up all these probabilities for $k = 1, ... , 4^{N_{bp}}$.
This is the Shannon entropy for an ensemble where all possible 
DNA sequences were realized.  If all humans would have the same
unique DNA sequence then the Shannon entropy of a
human and of the identical species would be
 \be
 \sigma_P^{bp} \ \ =\
H(X) = -p_i \ln p_i \ =\  1/4^{N_{bp}} \cdot \ln 4^{N_{bp}} \ 
=\  {N_{bp}}\cdot \ln 4 /4^{N_{bp}}\ =\ \ 1.386 {N_{bp}}/4^{N_{bp}}
\label{sigma-bp}
\ee
and thus for a single sequence without any variation in the species:
\be
\ln H(X) \ \ =\  \ln (4.483 \cdot 10^9) - N_{bp} \ln 4\
\ =\ \  - 4.483 \cdot 10^9
\ee

\subsection*{Single Nucleotide Polymorphisms}
\label{SNPs}

Every human is genetically different, and one of the most common
genetic variations is called Single Nucleotide Polymorphisms (SNP). 
An SNP could be the replacement of Cytosine with Adenine for example.
The two possible variations are two {\it alleles} for this base position.

Let us now estimate the number of different sequences in the 
whole human genome (i.e. in the total of 23 chromosomes or 23
DNAs. One person has only one DNA sequence in each of his or
her cells (except mutations developed due to external influences
in life). On the other hand different individuals in the
species may have variations in their DNA sequences. These
genetic variations are the Single Nucleotide Polymorphisms (SNP).
This happens on average once for 300 locations, 
and mostly only one type of exchange
is possible, i.e. one {\it allele}.  

Also, if we do not consider any other variation, the Shannon 
entropy can be calculated from eq. (\ref{HX}). 
We assume that the "standard" Human genome sequence is known.

We want to sum up the argument of
the Shannon entropy expression, $p_i \ln p_i$, over
all base pair configurations. We use the relation
for the summation 
$$
\sum\limits_{i=0}^{2^{N_{bp}}}\ \  \rightarrow \ \ \sum\limits_{i=0}^{{N_{bp}}} 
\frac{{N_{bp}}!}{({N_{bp}}{-}i)!\, i!}
$$

Shannon entropy for a sequence with $N_{bp}$ base pairs and $i$ SNP's is then:
\be
	H(X) \ \  =
	- \sum\limits_{i=0}^{{N_{bp}}} 
	\frac{{N_{bp}}!}{({N_{bp}}-i)!\,i!}P_A^i(1-P_A)^{{N_{bp}}{-}i} 
	\ln\left[P_A^i(1-P_A)^{N_{bp}{-}i}\right] \ ,
\ee
where  $P_A = 1/300$. This counts up the number of possible SNP 
changes that may happen.

Notice that \cite{MSSWF16}:
\be
\sum\limits_{i=0}^{N_{bp}} 
\frac{N_{bp}!}{(N_{bp}-i)!\,i!}P_A^i(1-P_A)^{N_{bp}{-}i} = 1.
\ee
The  function $\ln$ can be rewritten
\be
\ln\left[P_A^i(1{-}P_A)^{N_{bp}{-}i}\right]=
N_{bp}\, \ln\left[1{-}P_A\right]+i\, \ln\left[\frac{P_A}{1{-}P_A}\right] \ .
\ee
Putting this in the above equation we get
\be
	H(X) \ \ = - N_{bp} \ln\left[1{-}P_A\right] - 
	\ln\left[\frac{P_A}{1{-}P_A}\right] \times 
	\sum\limits_{i=0}^{N_{bp}}\left[\frac{N_{bp}!\ i}{(N_{bp}-i)!\,i!}
	\, P_A^i(1-P_A)^{N_{bp}{-}i} \right] \ .
\ee
Since the values of $i$ that contribute the most are around 
$N_{bp} \times P_A$ 
we use the approximation that the $i$ in the numerator 
is a constant instead of a variable.
\be
i \rightarrow k = N_{bp}P_A
\ee
We can then approximate $H(x)$
\be
	H(X) \ \  \approx - N_{bp} \ln \left[1-P_A\right] - 
	N_{bp} P_A \ln\left[\frac{P_A}{1-P_A}\right]\ = \ \
	N_{bp} \left[-P_A \ln\left[\frac{P_A}{1-P_A}\right]-\ln(1-P_A)\right] \ .
\ee
We see that $H(X)$ is proportional to $N_{bp}$  and $P_A = 1/300$ so that
%

%
\be
H_{max}(X)\ \ \approx \ 0.0223 \cdot N_{bp}\  = \ \ 7.15 \cdot 10^7 ,
\ee
for all possible SNP configurations of the Human species. This is 
much smaller than the entropy corresponding to the maximum
possibility of all variations for the DNA of a length of the
human genome.
\smallskip

The {\bf spatial configuration of the DNA} can lead to even higher 
complexity than the base-pair sequence. The spatial area that could
be reached by the DNA may reach $N_{bp}^3$, when the length is
$N_{bp}$. Thus the probability for one configuration could be of the 
order of $N_{bp}^{-2}$. On the other hand, our knowledge on the
spatial configuration of the DNA of the human genome is little,
and the DNA exists in different configurations: curled up in chromosomes,
or straightened out at cell division. Thus, to study the entropy arising
from the spatial configuration of DNA would require knowledge that
we do not have today.

Before cell division, the chromosomes are reshaped and the
DNA takes a linearly extended shape (like an extended line),
which enables more motion and the replication of the DNA.
In this configuration the degrees of freedom in the physical
phase space are increased, and thus the entropy of the
configuration also increases. This extending of the DNA molecule requires
additional energy temporarily. The linear shape allows more
motion and increases the possibility of occupying a larger
part of the phase space.

Thus the compact curled up configuration of DNA has smaller entropy,
having similar amount of freedom as atoms in a solid state, while 
the linearly extended molecule has considerably larger entropy,
similar to a liquid type of structure.  So, the entropy of the 
DNA changes dynamically.

\smallskip


The average {\bf weight of a DNA} can be obtained 
from its length, $N_{bp}=3.234\cdot 10^9$, times the number of atoms per base-pair, 650. Thus $m_{DNA} = 650 \cdot N_{bp} $ AMU $= 3.489 \cdot 10^{-15}$kg.
Consequently in 1 kg we have $N_{DNA} = 2.866 \cdot 10^{14}$ molecules.

Then, from eq. (\ref{sigma-bp}) we get the specific entropy for the DNA base-pair
sequence of a single individual. The corresponding entropy
in 1 kg of DNA molecules is then 
\be
S_{1kg}^{DN\!A} \ \ = \
k_B\, N_{DN\!A}\, \sigma_P^{DN\!A}\ = \ \  3.958\cdot10^{-1\,947\,000\,000} {\rm J/K} \ .
\nonumber
\label{SDNA}
\ee

\section{Entropy of Live Material Tissues}
\label{S5}

Live material tissues contribute to a high complexity on a 
larger scale. The best example of this is the nervous system.
In a human brain there are  
about $10^{10}$ nerve cells, and each of these have
about $10^3 \ - 10^5$  synaptic junctions. 
This means that the whole nervous system may have
up to $N_s=10^{15}$ synaptic junctions. 
If we quantize the synaptic couplings as 0 to 1 only, then we have  
$2^{N_s}$ different states statically.  This can even lead to higher
complexity than the DNAs.
  At the same time it is difficult to estimate, which configurations
  can be realized, and actually these configurations can change 
  each second.  A sleeping persons brain has different entropy than
  an active one.
  
To analyse this level of complexity is beyond our present goal.

\section{Conclusions}

We have demonstrated quantitatively the increasing complexity of materials,
and used the entropy for unit amount of material in order to be able to get
a measure. This idea stems from Ervin Schr\"odinger, but our
knowledge today makes it possible to extend the level of 
quantitative discussion to complex live materials.

We may continue these studies to higher levels of material structures,
like living species, artificial constructions, symbiotic coexistence of
different species, or groupings of the same species. We can even continue up to 
structures in Human society.

The main achievement of this work is to show how the entropy in the physical
phase space and the entropy of structural degrees of freedom (Shannon entropy) 
can be discussed on the same platform, as shown in section \ref{S2} via
eq. (\ref{ee14}).  For further developments it is important to point out two
fundamental aspects of the entropy concept: (i) the {\it quantization} of the 
space of a given degree of freedom, and (ii) the {\it selection of the realized, realizable
or beneficial configurations} from all the possible ones. If  we go towards 
more complex systems, these two questions become non-trivial,
and particularly in the case of the utmost complex systems where it is not clear
which are the realizable and most beneficial systems. This is already a 
challenging question in the case of the nervous system, and even more so
for organizations in the society or in economy.

The examples presented are all analyzed from a static point of view.
As we see on the example of the nervous system the dynamical 
change of the entropy of the system is also important. Furthermore ,
the speed of development is also important. The early development of
complexity happened on a very slow rate, while the development 
of complexity of the nervous system is many orders of magnitude more
rapid then the development of the DNA structure. The dynamics and
direction of these changes are also essential, as shown in ref. \cite{PCs1980}.

Some specific studies in these directions exist already. 
See e.g. considerations on general dynamics of sustainable development
\cite{Biro}, and 
on the life-span of different species related to their entropy and metabolism.
We are looking forward to further developments  \cite{PCs1980}.

\section*{Acknowledgements}

Enlightening discussions with 
Tam\'as Bir\'o, Stuart Holland, 
Zolt\'an N\'eda, G\'abor Pall\'o, Istv\'an Papp,
P\'eter V\'an and Yilong Xie 
are gratefully acknowledged.
This research was partially supported by the Academia Europaea Knowledge Hub Bergen, by the  Institute of Advanced Studies, K\H{o}szeg, and by the 
Research Council of Norway, grant no. 231469.


\end{document}